\documentclass[plb,reprint]{revtex4-1}
\usepackage{amssymb,amsfonts,amsmath,graphicx}

\usepackage{amssymb,amsmath,epsfig,todonotes,longtable}

\numberwithin{equation}{section}

\newcommand{\bean}{\begin{eqnarray*}}
\newcommand{\eean}{\end{eqnarray*}}

\newcommand{\fref}[1]{Figure~\ref{#1}}

\newcommand{\capt}[3]{\parbox{#1}{\renewcommand{\baselinestretch}{1.2}
                                                           \caption{\label{#2}\small\it #3}}}

\newcommand{\IP}{\mathbb{P}}

\newcommand{\IZ}{\mathbb{Z}}
\newcommand{\cO}{{\cal O}}
\newcommand{\cN}{{\cal N}}

\newcommand{\cA}{{\cal A}}
\newcommand{\cB}{{\cal B}}
\newcommand{\cC}{{\cal C}}

\newcommand{\cK}{{\cal K}}
\newcommand{\cV}{{\cal V}}

\def\cjn1{{\cA, \cC^*\otimes \wedge^j \cN^*}}
\def\bjn1{{\cA, \cB^*\otimes \wedge^j \cN^*}}
\def\vjn1{{\cA, \cV^*\otimes \wedge^j \cN^*}}
\def\cjn2{{\cA, \cC\otimes \wedge^j \cN^*}}
\def\bjn2{{\cA, \cB\otimes \wedge^j \cN^*}}
\def\vjn2{{\cA, \cV\otimes \wedge^j \cN^*}}

\newcommand{\cicy}[2]{\begin{matrix} #1\end{matrix}\!\left[\begin{matrix}#2 \end{matrix}\right]}

\newcommand{\floor}[1]{\left\lfloor{#1}\right\rfloor}
\newcommand{\ceil}[1]{\left\lceil{#1}\right\rceil}

\newcommand{\mc}{\mathcal}

\newcommand{\be}{\begin{equation}}
\newcommand{\ee}{\end{equation}}
\newcommand*{\nnbe}{\begin{equation}}
\newcommand*{\nnee}{\end{equation}}
\newcommand{\bea}{\begin{eqnarray}}
\newcommand{\eea}{\end{eqnarray}}
\newcommand{\ba}{\begin{align}}
\newcommand{\ea}{\end{align}}
\newcommand{\bi}{\begin{itemize}}
\newcommand{\ei}{\end{itemize}}

\newsavebox{\overlongequation}

\begin{document}

\title{Flops, Gromov-Witten Invariants  and \\ Symmetries of Line Bundle Cohomology on \\ Calabi-Yau Three-folds}



\author{Callum R. Brodie}
\email[]{callum.brodie@ipht.fr}
\affiliation{Institut de Physique Th\'{e}orique, Universit\'{e} Paris Saclay, CEA, CNRS \\ Orme des Merisiers, 91191 Gif-sur-Yvette CEDEX, France}

\author{Andrei Constantin}
\email[]{andrei.constantin@physics.ox.ac.uk}
\affiliation{Rudolf Peierls Centre for Theoretical Physics, University of Oxford, Parks Road, Oxford OX1 3PU, UK}
\affiliation{Mansfield College, University of Oxford, Mansfield Road, OX1 3TF, UK}

\author{Andre Lukas}
\email[]{andre.lukas@physics.ox.ac.uk}
\affiliation{Rudolf Peierls Centre for Theoretical Physics, University of Oxford, Parks Road, Oxford OX1 3PU, UK}

\vspace{12pt}

\begin{abstract}\noindent
The zeroth line bundle cohomology on Calabi-Yau three-folds encodes information about the existence of flop transitions and the genus zero Gromov-Witten invariants. We illustrate this claim by studying several Picard number $2$ Calabi-Yau three-folds realised as complete intersections in products of projective spaces. Many of these manifolds exhibit certain symmetries on the Picard lattice which preserve the zeroth cohomology.  \end{abstract}

\maketitle		

\section{Introduction}
It has long been known that the understanding of the massless states resulting from string compactifications relies on the understanding of certain cohomology groups on the internal space. To achieve this, one often has to go through lengthy computations of bundle-valued cohomology groups on Calabi-Yau three-folds. Various algebraic and topological tools can be employed to derive cohomology from local data, but such methods inevitably hide away much of the information encoded therein. The purpose of the present letter is to highlight the richness of structure present in the zeroth cohomology of line bundles on Calabi-Yau three-folds by case-studying a small number of Picard number two manifolds realised as complete intersections in products of projective spaces (CICY three-folds)~\cite{Candelas:1987kf,Green:1987cr}. In particular, in these examples one can discern the presence of flops, the value of certain genus zero Gromov-Witten invariants, and the existence of symmetries in the cohomology data. 

This study is part of the greater quest to understand the extent to which line bundle and more generally vector bundle cohomology can be expressed in terms of analytic formulae on spaces of interest in string theory. A systematic understanding of these formulae, for which we present the first steps in this work, will likely open up new approaches for bottom-up string model building. Initial evidence for the existence of such formulae has been obtained through a combination of direct observation \cite{Constantin:2018otr, Buchbinder:2013dna, Constantin:2018hvl, Larfors:2019sie, Brodie:2019pnz} and machine learning \cite{Klaewer:2018sfl, Brodie:2019dfx} of line bundle cohomology dimensions computed algorithmically on several classes of two and three-dimensional complex manifolds, such as complete intersections in products of projective spaces, toric varieties and hypersurfaces therein, del Pezzo and Hirzebruch surfaces. Subsequently, for certain classes of  surfaces widely used in string theory such as toric surfaces, weak Fano surfaces and K3 surfaces, explicit formulae describing all cohomology groups of line bundles have been established through rigorous proofs \cite{Brodie:2019ozt, Brodie:2020wkd}.

Much less is currently understood about the structure of bundle cohomology on three-folds, except in the case of simple elliptic fibrations over two-dimensional bases \cite{Brodie:2020wkd} or for certain divisors on toric hypersurfaces~\cite{Braun:2017nhi}. The empirical evidence suggests that the Picard group can be divided into disjoint regions, in each of which the cohomology dimensions are described by functions that are polynomial or very close to polynomial in the first Chern class of the line bundle. This seems to be the case for the zeroth as well as for all higher cohomologies, though here we will only study the zeroth cohomology.  

\section{Generalities}

To set the scene, let $X\subset {\cal A}$ be a smooth CICY three-fold defined as the common zero locus of several multi-homogeneous polynomials in the coordinates of the product space  ${\cal A}=\IP^{n_1}\times\dots\times\IP^{n_m}$. The multi-degrees of the defining polynomials can be recorded as the columns of a matrix, known as the configuration matrix, of the form
\begin{equation}
\label{conf}
\cicy{\IP^{n_1}\\[4pt] \vdots\\[4pt] \IP^{n_m}}{q^1_1&\cdots&q^1_R\\[4pt] \vdots&\cdots&\vdots\\[4pt]q^m_1&\hdots&q^m_R}^{h^{1,1}(X),~h^{2,1}(X)}
\end{equation}
where $h^{1,1}(X)$ and $h^{2,1}(X)$ are the two non-trivial Hodge numbers of $X$. The condition that $X$ has vanishing first Chern class corresponds to the condition that the sum of the degrees in each row of the configuration matrix equals the dimension of the corresponding projective space plus one. All CICY three-folds are simply connected. 

Holomorphic line bundles are specified by their first Chern class, which is an element of $H^2(X,\mathbb Z)$. All Calabi-Yau manifolds $X\subset\cA$ discussed in this letter benefit from being `favourably' embedded, in the sense that a basis of $H^2(X,\mathbb Z)$ can be obtained by pulling-back the K\"ahler two-forms of the hyperplane bundles over the $\IP^n$ factors of $\cA$. We denote this basis by $(D_1,D_2,\ldots, D_m)$ for ${\cal A}=\IP^{n_1}\times\dots\times\IP^{n_m}$ and the dual basis of curve classes by $(C_1,C_2,\ldots, C_m)$. The vast majority of CICY three-folds can be favourably embedded  \cite{Anderson:2017aux}. 

Several cones in $H^2(X,\mathbb R)$ play an important role in the present discussion, whose definition we briefly review. The {\itshape K\"ahler cone} $\cK$ is the set of cohomology classes of smooth positive definite closed $(1, 1)$-forms. For all the manifolds studied below, the K\"ahler cone descends from the ambient product of projective spaces, which means $\cK$ is the positive span of $\{D_1,D_2,\ldots, D_m\}$. The closure $\overline \cK$ is the {\itshape nef cone}. 
A line bundle is nef if its first Chern class belongs to the nef cone. A line bundle is called {\itshape effective} if it has a global section or, equivalently, a non-vanishing zeroth cohomology. 

If $L$ is a line bundle in the interior of the nef cone, Kodaira's vanishing theorem guarantees that all higher cohomologies are trivial and, consequently, $h^0(X,L) = {\chi}(X,L)$, where ${\chi}(X,L)$ is the Euler characteristic of $L$, which on a Calabi-Yau three-fold takes form
\begin{equation}\label{eq:index}
{\chi}(X,\cO_X(D)) = \frac{1}{6}D^3 + \frac{1}{12} c_2(X)\cdot D~.
\end{equation}

The Euler characteristic is a linear combination of two basic topological invariants on $H^2(X,\mathbb Z)$, namely the cup-product cubic form $H^2(X,\mathbb Z)\rightarrow \mathbb Z$ given by $D\mapsto D^3$ and the linear form $c_2:H^2(X,\mathbb Z)\rightarrow \IZ$ given by $D\mapsto c_2(X)\cdot D$. It is known that a nef line bundle $L=\cO_X(D)$ on a projective three-fold falls in the interior of the effective cone iff $D^3>0$ (see Thm.~2.2.16. in~\cite{lazarsfeld2004positivity1}). 

We will make use of three important relations that hold when $X$ and $X'$ are related by a flop which contracts a finite number of disjoint $\IP^1$ curves. First, since a flop is an isomorphism in co-dimension one, $H^2(X,\mathbb R)$ and $H^2(X',\mathbb R)$ can be identified. 
In the following, divisors identified in this way will be denoted by the same symbol, primed for $X'$ and unprimed for $X$.
Second, since the zeroth cohomology counts co-dimension one objects, it is preserved under the flop, that is, $h^0\big(X, \mc{O}_X(D)\big) = h^0\big(X', \mc{O}_{X'}(D')\big)$ where $D'$ is the divisor on $X'$ corresponding to $D$ on $X$. Note the same is not true of higher cohomologies. Third, the above two forms have the following transformation rule,
\begin{equation}
\begin{aligned}
D'^3 &= D^3 - \sum_i (D\cdot \cC_i)^3\\[-4pt]
c_2(X')\cdot D' &= c_2(X)\cdot D +2 \sum_i D\cdot \cC_i~,
\end{aligned}
\end{equation}
where $\cC_1,\cC_2,\ldots,\cC_N$ are the isolated exceptional $\IP^1$ curves with normal bundle $\cO(-1)\oplus\cO(-1)$ contracted in the flop \cite{friedman_1991, wilson_1999}. 
The K\"ahler cones $\cK(X)$ and $\cK(X')$ share a common wall. 
The change in the cup product cubic form corresponds in topological string theory to the statement that the A-model 3-point correlation function on $\cK(X)$ may be analytically continued to give the A-model 3-point correlation function on $\cK(X')$ \cite{morrison_1996}.

\section{The manifold $7887$.} 
In this example $X$ is a generic Calabi-Yau hypersurface in the ambient space ${\cal A}=\mathbb{P}^1\times\mathbb{P}^3$ defined by the configuration matrix 
\begin{equation}
\cicy{\IP^1 \\ \IP^3}{\,2~ \\ \,4~}^{2,86}
\end{equation}
with identification number $7887$ in the CICY list~\cite{Candelas:1987kf,Green:1987cr}.

If $L$ is a line bundle over $X$, we write its first Chern class as $c_1(L)=k_1 D_1 + k_2 D_2$. Line bundle cohomology dimensions, computed algorithmically using the CICY package \cite{CICYpackage} for $-3\leq k_1\leq 4$, $-1\leq k_2\leq 9$ are shown in the chart below. 

\begin{figure}[h]
\begin{center}
\includegraphics[width=7.8cm]{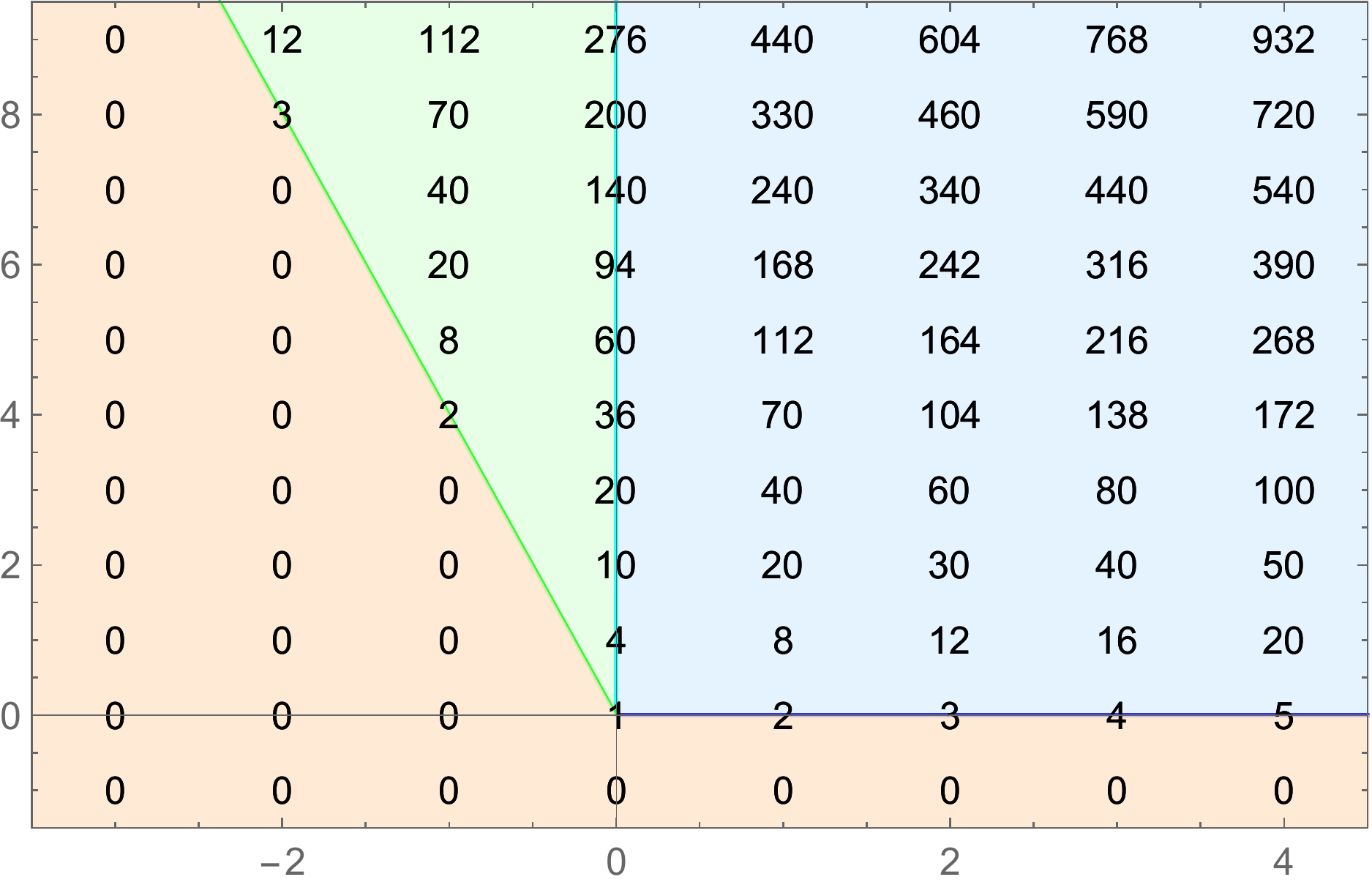}
\capt{3.2in}{fig:X7887}{Zeroth cohomology dimensions on the CICY manifold 7887. Blue region:   K\"ahler cone ${\cal K}(X)$ of $X$. Green region: K\"ahler cone ${\cal K}(X')$ of the flopped space~$X'$.}
\vspace{-12pt}
\end{center}
\end{figure}

The positive quadrant (blue region in Fig.~\ref{fig:X7887}) corresponds to the K\"ahler cone of $X$. In this region we have $h^0(X,L) = \chi(X,L)$. The Euler characteristic is computed with the following topological data: 
\begin{equation}\label{isec7887}
\begin{array}{cccccc}
d_{111}&d_{112}&d_{122}& d_{222} & c_2\cdot D_1 & c_2 \cdot D_2\\\hline
0 & 0 & 4 & 2 & 24 & 44
\end{array}
\end{equation}
where $d_{ijk} = D_i\cdot D_j\cdot D_k$. Along the horizontal boundary of the nef cone the cubic form $D\mapsto D^3$ vanishes, which indicates that this is also a boundary of the effective cone. The vertical boundary is shared with another cone (the green region in Fig.~\ref{fig:X7887}) which we conjecture to be the nef cone of a flopped Calabi-Yau three-fold $X'$.  For line bundles $L$ in this region this implies
\begin{equation}
h^0(X,L) = h^0(X',L') = \chi(X',L') \,.
\end{equation}
Indeed, fitting the cohomology data to the formula \eqref{eq:index}, one finds the following topological data for $X'$
\begin{equation}\label{isec7887prime}
\begin{array}{cccccc}
d_{111}'&d_{112}'&d_{122}'& d_{222}'& c_2'\cdot D_1' & c_2' \cdot D_2'\\\hline
-64 & 0 & 4 & 2 & 152 & 44
\end{array}
\end{equation}
where $D_i'$ is the divisor on $X'$ corresponding to the divisor $D_i$ on $X$. These changes are consistent with the hypothesis that $X$ and $X'$ are related by a flop in which $64$ isolated $\IP^1$ curves with class $C_1$ are being contracted. 

In fact, this is precisely the number of genus zero curves in the class $C_1$, which can be interpreted as the corresponding Gromov-Witten invariant. It is easy to count these. Denoting the coordinates of the ambient space by $(x_a,y_b)$, with $a=0,1$ and $b=0,1,2,3$, the defining equation takes the form $x_0^2 P(y_b)+x_0x_1Q(y_b)+x_1^2 R(y_b)=0$. When $P(y_b)=Q(y_b)=R(y_b)=0$, there is an entire $\IP^1$ worth of solution and, since $P, Q, R$ have degree $4$ this happens at precisely $4^3 = 64$ points in $\IP^3$. 

The curve class $C_1$ is orthogonal to the wall separating the K\"ahler cones $\cK(X)$ and $\cK(X')$, which together form what is known in the Physics literature as the {\itshape extended K\"ahler cone} or,  in the Mathematics literature, as the {\itshape movable cone}. The other boundary of $\cK(X')$ corresponds to $k_2=-4k_1$, with $k_1\leq 0$, and along this edge the cup-product cubic form vanishes, indicating that this is also a boundary of the effective cone. This provides evidence in support of the claim that $X$ has only two birational minimal models related by a flop. In particular, $X$ is an example of a Mori dream space \cite{hu2000}. The boundaries of the effective cone are at finite distance, as measured with the moduli space metric, but it is evident from the intersection numbers in Eqs.~\eqref{isec7887} and \eqref{isec7887prime} that the volume (proportional to $d_{ijk}t^it^jt^k$ where $t^i$ are the K\"ahler moduli) vanishes at the boundaries of the effective cone. 

\begin{figure}[h]
\begin{center}
\includegraphics[width=7.8cm]{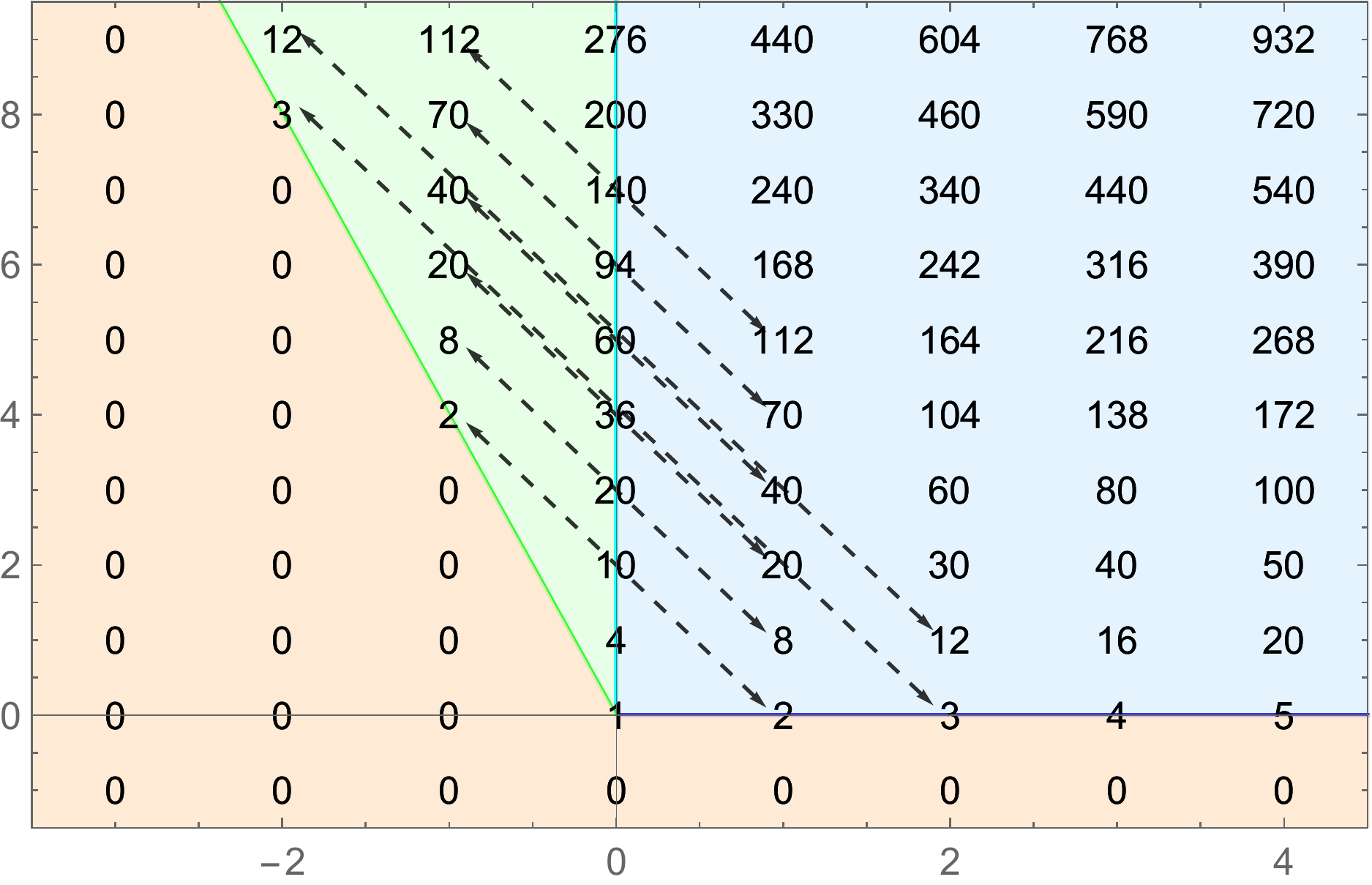}
\capt{3.2in}{fig:X7887symm}{A $\IZ_2$ symmetry of the zeroth coh.~for CICY $7887$.}
\vspace{-12pt}
\end{center}
\end{figure}

What can be said about the flopped manifold $X'$? On general grounds, we know that it is a smooth Calabi-Yau three-fold with the same Hodge numbers as $X$. We also know the triple intersection numbers and the $c_2$ form, written in the basis $\{D_1',D_2'\}$, as given in Eq.~\eqref{isec7887prime}. Choosing as a basis of $H^2(X',\IZ)$ the generators of the K\"ahler cone $\cK(X')$, namely $\tilde D_1 = -D_1' + 4 D_2'$ and $\tilde D_2 = D_2'$, the triple intersection numbers and the $c_2$ form become identical to those of $X$. As such, $X'$ and $X$ are diffeomorphic to each other \cite{WALL1966, wilson2017}. 

It is not surprising then that the zeroth cohomology displays a $\IZ_2$ symmetry
\begin{equation}\label{7887symmetry}
h^0(X, \cO_X({\bf k}))=h^0(X, \cO_X(M {\bf k})) 
\end{equation}
with generator\vspace{-12pt}
\begin{equation} 
M = \left(\begin{array}{cc} \!\!-1 & 0 \\ ~n & 1 \end{array}\right)\; , \label{Z2gen}
\end{equation}
where ${\bf k} = (k_1, k_2)^T$ and $n=4$  (see Fig.~\ref{fig:X7887symm}). This follows only for the zeroth cohomology, as these are preserved under a flop.

Finally, we write down explicit topological formulae for the zeroth cohomology of line bundles on $X$. Inside $\cK(X)$, Kodaira's vanishing theorem guarantees that $h^0(X,L)=\chi(X,L)$. Similarly, inside $\cK(X')$ we have $h^0(X,L)=h^0(X',L')=\chi(X',L')$ where $L'$ is the line bundle on $X'$ corresponding to $L$ on $X$. The wall between the two K\"ahler cones falls in the interior of the effective cone and in this case the Kawamata-Viehweg vanishing theorem implies  $h^0(X,L)=\chi(X,L)=\chi(X',L')$. For the trivial bundle $h^0(X,\cO_X)=1$, since $X$ has a single connected component. For line bundles $L=\cO_X(k_1 D_1)$ lying along the edge $k_1> 0$, we have $h^0(X,L)=\chi(\IP^1, k_1 H_{\IP^1})$, where $H_{\IP^1}$ is the hyperplane class in $\IP^1$, a result which can be traced back by sequence chasing. The $\IZ_2$ symmetry \eqref{7887symmetry} implies then that a similar relation must hold along the other boundary of the effective cone, $h^0(X,\cO_X(k_1D_1-4k_1 D_2))=\chi(\IP^1, -k_1 H_{\IP^1})$, for $k_1<0$. We sum up these formulae in the following table: 
\begin{equation*}
\begin{tabular}{ l | c}
 {\rm region in eff.~cone}		&~$ h^0(X,L=\cO_X(k_1D_1+k_2D_2))$ \\
\hline
$\cK(X)$ &~ $\chi(X,L)$ \\
$\cK(X')$ &~ $\chi(X',L')$ \\
$k_1=0,~k_2>0$ &~ $\chi(X,L)=\chi(X',L')$ \\
$k_1<0,~k_2=-4k_1~$ & $\chi(\IP^1,-k_1H_{\IP^1})$\\
$k_1>0,~k_2=0$ & $\chi(\IP^1,k_1H_{\IP^1})$\\
$k_1=k_2=0$ & $1$
\end{tabular}
\label{eq:7887_formulae}
\end{equation*}
Similar results, with varying values of $n$ in Eq.~\eqref{Z2gen}, are obtained for $19$ other Picard number $2$ three-folds in the CICY list, namely those with identifiers 7643, 7668, 7725, 7758, 7806, 7808, 7816, 7819, 7822, 7823, 7833,
7844, 7853, 7867, 7869, 7882, 7883, 7886 and 7888.

\section{The manifold $7885$.} 
In this section $X$ is a generic smooth Calabi-Yau three-fold belonging to the family described by the configuration matrix
\vspace{-12pt}
\begin{equation}
\cicy{\IP^1 \\ \IP^4}{\ 1~&1~ \\ \,4~&1~}^{2,86}
\end{equation}
with identifier $7885$ in the CICY list.
\begin{figure}[h]
\begin{center}
\includegraphics[width=7.8cm]{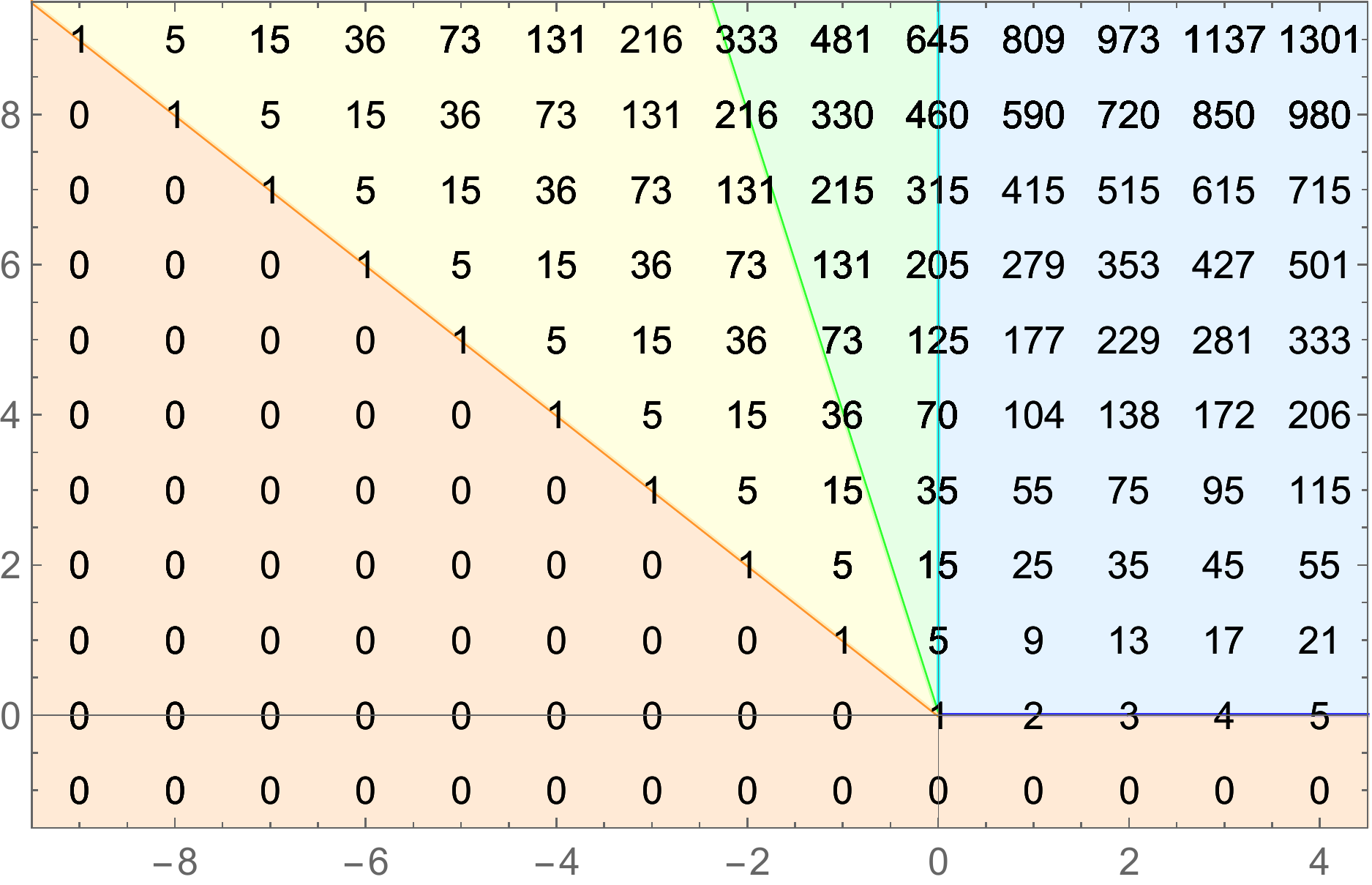}
\capt{3.3in}{fig:X7885}{Zeroth cohomology dimensions on 7885. In blue: $\cK(X)$. In green: $\cK'(X')$. In yellow: the Zariski chamber $\Sigma$.}
\vspace{-12pt}
\end{center}
\end{figure}

As before, the positive quadrant corresponds to the K\"ahler cone $\cK(X)$ where $h^0(X,L) = \chi(X,L)$ and the Euler characteristic is computed with the topological data: 
\begin{equation}
\begin{array}{cccccc}
d_{111}&d_{112}&d_{122}& d_{222} & c_2\cdot D_1 & c_2 \cdot D_2\\\hline
0 & 0 & 4 & 5 & 24 & 50
\end{array}
\end{equation}
Along the horizontal boundary of the nef cone $D^3=0$, hence this is also a boundary of the effective cone. The other boundary corresponds to a wall separating $\cK(X)$ from what we conjecture to be the K\"ahler cone of a flopped manifold $X'$. Fitting the zeroth cohomology data to the Euler characteristic formula we find the following topological data for $X'$
\begin{equation}\label{isec7885prime}
\begin{array}{cccccc}
d_{111}'&d_{112}'&d_{122}'& d_{222}'& c_2'\cdot D_1' & c_2' \cdot D_2'\\\hline
-16 & 0 & 4 & 5 & 56 & 50
\end{array}
\end{equation}
The changes in the triple intersection numbers and the $c_2$ form are consistent with the hypothesis that $X$ and $X'$ are related by a flop in which $16$ isolated $\IP^1$ curves with class $C_1$ are being contracted. This is indeed the genus zero Gromov-Witten invariant in the class $C_1$. 

The left boundary of $\cK(X')$ is the edge $k_2=-4k_1$, with $k_1\leq 0$. The cup product cubic form does not vanish along this edge, which must then be contained inside the effective cone. This is consistent with the cohomology data shown in \fref{fig:X7885} which indicates the presence of a third cone inside the effective cone, denoted by~$\Sigma$. 

The K\"ahler cones $\cK(X)$ and $\cK(X')$ together form the movable cone. It is known that effective integral divisor classes lying outside of the movable cone have fixed components. This is consistent with the presence of a rigid divisor class, namely $\Gamma=-D_1+D_2$, which can be inferred from $h^0(X,\cO_X(\Gamma))=h^0(X',\cO_{X'}(\Gamma'))=1$. This rigid divisor is part of the fixed locus of every linear system in the cone $\Sigma$. The presence of $\Gamma$ and the amount by which it is present can be detected by intersection properties leading to a Zariski decomposition of every divisor class in $\Sigma$ in the sense of \cite{Nakayama_2004}. We will not attempt to say anything general about the existence of the Zariski decomposition on three-folds which is a difficult problem in itself. However, the cohomology data in the yellow region of Fig.~\ref{fig:X7885} shows an obvious pattern - it is constant along the diagonals - which is consistent with the following picture. 

Let $\tilde D_1'=-D_1' + 4 D_2'$ and $\tilde D_2' = D_2'$ denote the generators of the nef cone $\overline\cK(X')$ and let $\tilde C_1'$ and $\tilde C_2'$ denote the two dual curve classes, $\tilde C_i'\cdot \tilde D_j' = \delta_{ij}$, where the intersection is computed with the data \eqref{isec7885prime} for $X'$. Let $D'$ be an effective divisor class on $X'$ and assume it has a Zariski decomposition $D' = P' + N'$, where $P'$ and $N'$ are $\mathbb Q$-divisor classes, $N'$ is effective and $P'$ is nef, that is $P'\in \overline\cK'(X')$. 

From now on we assume that $D'\in \Sigma$, which is the case we are interested in. In this case, $P'$ lies on the boundary shared by $\Sigma$ and $\overline\cK(X')$. From general properties of Zariski decomposition we know that
\begin{equation}
h^0(X',\cO_{X'}(D')) = h^0(X', \cO_{X'}(\floor{P'}))~,
\end{equation}
where $\floor{P'}$ is the round down version of $P'$. The round down version of a $\mathbb Q$-divisor is defined as the divisor obtained by rounding down all the coefficients in its expansion. In the context of Zariski decomposition, when $D'$ is an integral divisor class, the class $\floor{P'}$ is well-defined, as explained in \cite{Brodie:2020wkd}. 

The divisor class $N'$ is a rational multiple $N' = \gamma \Gamma'$ of the rigid divisor class $\Gamma'=\tilde D_1'-3\tilde D_2'$ and $P'$ is a rational multiple of~$\tilde D_1'$. Since $D'= P'+N'$ and $\tilde D_1'\cdot \tilde C_2'=0$, it follows that $D'\cdot \tilde C_2' = N'\cdot\tilde C_2'= \gamma \Gamma' \cdot \tilde C_2'$, from which $\gamma = D'\cdot \tilde C_2'/ \Gamma' \cdot \tilde C_2'$. Consequently, 
\begin{equation}
\begin{aligned}
h^0(X',~&\cO_{X'}(D')) = h^0(X', \cO_{X'}(D'{-}\ceil{\gamma}\Gamma'))~\\
& = \chi\left(X', \cO_{X'}\left(D'-\ceil{\frac{D'\cdot\tilde C_2'}{\Gamma'\cdot\tilde C_2'}}\Gamma'\right)\right),
\end{aligned}
\end{equation}
which is also equal to $h^0(X,\cO_X(D))$, where $D$ is the divisor class on $X$ corresponding to $D'$ on $X'$. With $D\ = k_1D_1+k_2 D_2$ it is straightforward to compute $\gamma = (4k_1+k_2)/(-3)$.

We summarise the zeroth cohomology formulae on the manifold $7885$ in the following table: 
\begin{equation*}
\begin{tabular}{ l | c}
 {\rm region in eff.~cone}		&~$ h^0(X,L=\cO_X(D=k_1D_1+k_2D_2))$ \\
\hline
$\cK(X)$ &~ $\chi(X,\cO_X(D))$ \\[3pt]
$\overline\cK(X') \setminus \{\mc{O}_{X}\}$ &~ $\chi(X',\cO_{X'}(D')$ \\[3pt]
$\overline\Sigma$ & ~ $\chi\left(X', \cO_{X'}\left(D'-\ceil{\frac{D'\cdot\tilde C_2'}{\Gamma'\cdot\tilde C_2'}}\Gamma'\right)\right)$ \\[4pt]
$k_1>0,~k_2=0$ & $\chi(\IP^1,(D\cdot C_1)H_{\IP^1})$\\[3pt]
$k_1=k_2=0$ & $1$
\end{tabular}
\label{eq:7885_formulae}
\end{equation*}
Seven other Picard number 2 CICY three-folds, with the identifiers 7807, 7817, 7840, 7858, 7868, 7873 and 7883, can be treated in a similar way.

\section{The manifold $7863$.} 
In this section $X$ is a generic smooth Calabi-Yau three-fold belonging to the family described by the configuration matrix 
\begin{equation}
\cicy{\IP^3 \\ \IP^3}{\ 2~&1~&1~ \\\ 2~&1~&1~ }^{2,66}
\end{equation}
with identifier $7863$ in the CICY list.

The manifold has an obvious (non-freely acting) $\IZ_2$ symmetry inherited from an ambient space symmetry that exchanges the two $\IP^3$ spaces. This symmetry is evident in the cohomology data shown in \fref{fig:X7863}.

\begin{figure}[h]
\begin{center}
\includegraphics[width=7.8cm]{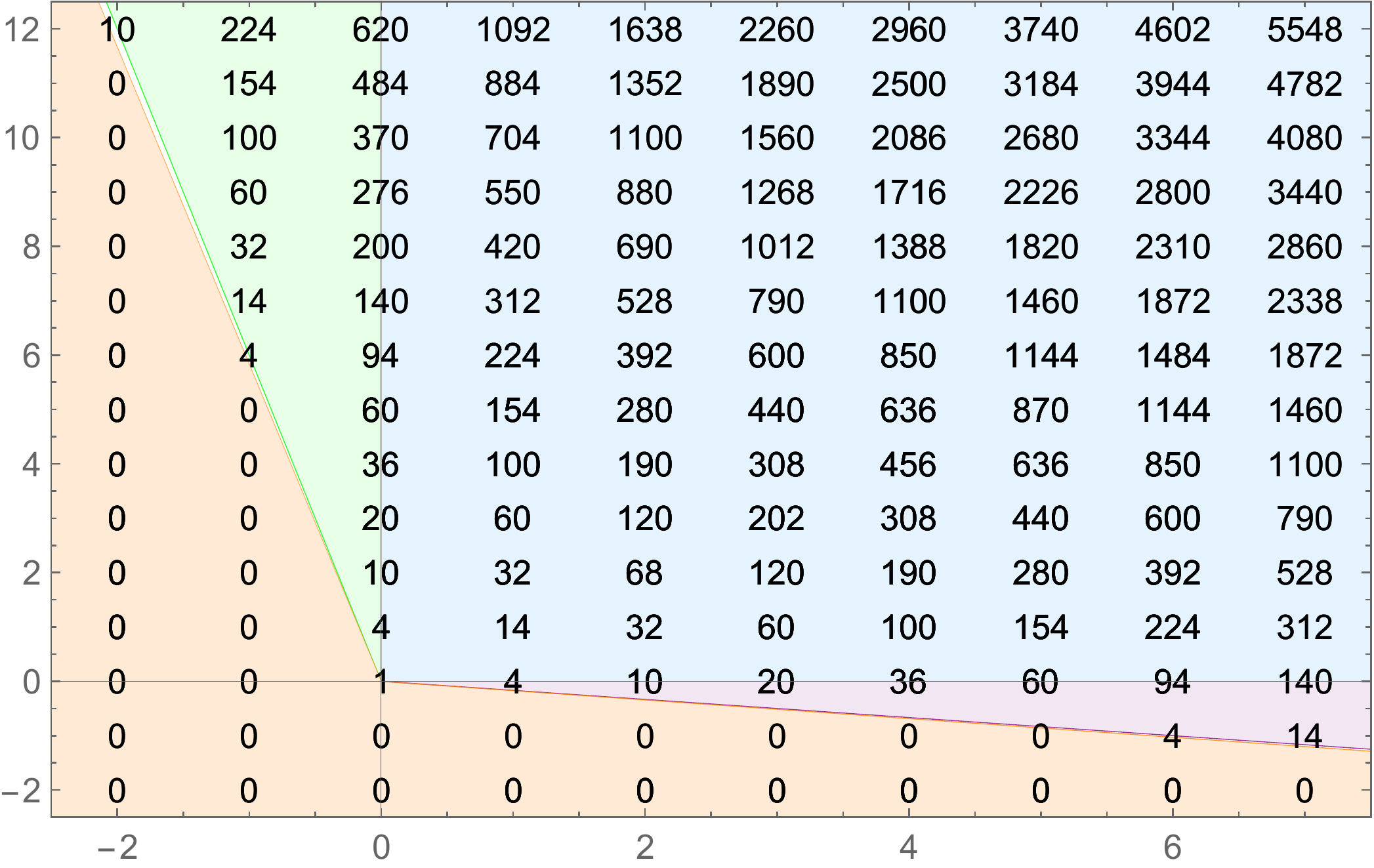}
\capt{2.9in}{fig:X7863}{Zeroth cohomology dimensions for the CICY manifold 7863.}
\vspace{-12pt}
\end{center}
\end{figure}

In the K\"ahler cone $\cK(X)$ (the blue region in Fig.~\ref{fig:X7863}), we have $h^0(X, L) = \chi(X,L)$ and the index is computed with the following topological data: 
\begin{equation}\label{7863topdata}
\begin{array}{cccccc}
d_{111}&d_{112}&d_{122}& d_{222} & c_2\cdot D_1 & c_2 \cdot D_2\\\hline
2 & 6 & 6 & 2 & 44 & 44
\end{array}
\end{equation}
In this example the cup-product cubic form does not vanish along either of the two boundaries of the nef cone $\overline\cK(X)$, hence neither of these is a boundary of the effective cone. There are two cones neighbouring $\cK(X)$, which we denote by $\cK(X'^{(1)})$ and $\cK(X''^{(1)})$, with generators $\{(-1,6),(0,1)\}$ and $\{(6,-1),(1,0)\}$, respectively. The reasons for using this slightly cumbersome notation will become evident in the following. As before, we conjecture that these regions are the K\"ahler cones of two flopped Calabi-Yau three-folds $X'^{(1)}$ and $X''^{(1)}$. Fitting the zeroth cohomology data to the Euler characteristic in these cones, the following topological data is obtained for $X'^{(1)}$ and $X''^{(1)}$:
\begin{align}
\label{isec7863prime}
\begin{array}{cccccc}
d_{111}'&d_{112}'&d_{122}'& d_{222}'& c_2'\cdot D_1' & c_2' \cdot D_2'\\\hline
-110 & 6 & 6 & 2 & 220 & 44
\end{array}~\\
\begin{array}{cccccc}
d_{111}''&d_{112}''&d_{122}''& d_{222}''& c_2''\cdot D_1'' & c_2'' \cdot D_2''\\\hline
2 & 6 & 6 & -110 & 44 & 220 
\end{array}
\end{align}
Due to the underlying $\IZ_2$ symmetry of the cohomology data, it is sufficient to focus on the manifold $X'^{(1)}$. 

The changes in the topological numbers \eqref{isec7863prime} are consistent with a flop between $X$ and $X'^{(1)}$ in which a total of $84$ isolated curves with normal bundle $\cO(-1)\oplus\cO(-1)$ are contracted, $80$ of which are in the numerical class $C_1$ and $4$ of which are in the numerical class $2C_1$. The number $80$ does indeed match the genus zero Gromov-Witten invariant in class $C_1$.

It is useful to recast the topological data \eqref{isec7863prime} in a basis which uses the generators of the K\"ahler cone $\cK(X'^{(1)})$, namely $\tilde D_1' = -D_1'+6D_2'$ and $\tilde D_2' = D_2'$. In this basis, the cup product cubic form and the $c_2$ form become identical to those on $X$, which indicates that $X$ and $X'^{(1)}$ are diffeomorphic. As such the cohomology data must be invariant under the $\IZ_2$ symmetry that exchanges $D_1$ and $\tilde D_1$ and fixes $D_2=\tilde D_2$, which is
\begin{equation}\label{7863symmetry}
h^0(X, \cO_X({\bf k})){=}h^0(X, \cO_X(M_1 {\bf k}))~,~M_1{=}\bigg(\begin{array}{cc} \!\!\!\!{-}{1} & \!\!{0}\!\! \\ ~\!\!{n}_{1} & \!\!{1}\!\! \end{array}\bigg)
\vspace{2pt}
\end{equation}
with $n_1=6$, mapping $\cK(X)$ to $\cK(X'^{(1)})$. Similarly, there is an involution
\begin{equation}\label{7863symmetry2}
h^0(X, \cO_X({\bf k})){=}h^0(X, \cO_X(M_2 {\bf k}))~,~M_2{=}\bigg(\begin{array}{cc} \!\!\!{-}{1} & \!\!{n_{2}}\!\! \\ ~\!0 & \!\!{1}\!\! \end{array}\bigg)
\vspace{2pt}
\end{equation}
with $n_2=6$, mapping $\cK(X)$ to $\cK(X''^{(1)})$. 

The existence of these involutions has important consequences. Under $M_1$, the K\"ahler cone $\cK(X''^{(1)})$ must be mapped to a new cone, lying to the left of $\cK(X'^{(1)})$ with generators $\{(-1,6),(-6,35)\}$, which we denote by $\cK(X'^{(2)})$. 
Similarly, the image of $\cK(X'^{(1)})$ under $M_2$ will be a new cone, lying below $\cK(X''^{(1)})$, which we denote by $\cK(X'^{(2)})$. These cones are very sharp and were not represented in Fig.~\ref{fig:X7863}. 

The two involutions do not commute. Together, by acting on $\cK(X)$ they generate two infinite series of K\"ahler cones $\cK(X'^{(i)})$ and $\cK(X''^{(i)})$, whose envelope is the (irrational) cone with generators $(-1,3+2\sqrt 2)$ and $(1,-3-2\sqrt 2)$, which we conjecture to be the pseudo-effective cone of divisors. Each of the manifolds $X'^{(i)}$ and $X''^{(i)}$ is a birational model of $X$, diffeomorphic to~$X$. This provides an example of a variety with an infinite number of Mori chambers. 

In each of these K\"ahler cones, the zeroth line bundle cohomology can be computed as an index, for example $h^0(X,\cO_X(D))=\chi(X'^{(i)},\cO_{X'^{(i)}}(D'^{(i)}))$, for $D \in\cK(X'^{(i)})$. The Euler characteristic can be easily computed, since in a basis of generators of $\cK(X'^{(i)})$ the required topological data is identical with that of $X$ given in \eqref{7863topdata} and the generators can be found iteratively by applying the two involutions. Similar statements hold for $D\in\cK(X''^{(i)})$.

On each of the manifolds $X'^{(i)}$ the boundary of the K\"ahler cone is at a finite distance in moduli space. In fact, since the intersection form remains unchanged, these distances are the same for all $X'^{(i)}$. This means that the two boundaries of the effective cone are at an infinite distance in moduli space.  

Several other CICY manifolds with Picard number $2$, including those with indentifiers 7644, 7726, 7759, 7761 and 7799  display similar features, with varying values of the numbers $n_1$ and $n_2$ in Eqs~\eqref{7863symmetry} and \eqref{7863symmetry2}. For some of these $n_1\neq n_2$, and there is no $\IZ_2$ symmetry inherited from the ambient space.

\section{Conclusions}
The main lesson to learn from the present work is that the zeroth line bundle cohomology on a Calabi-Yau three-fold $X$ encodes a wealth of information about the flops connecting the birational models of $X$, as well as about certain Gromov-Witten invariants. This insight can facilitate the computation of the true effective cone, which in some cases is irrational. In all the cases studied here we were able to decompose the effective cone into cohomology chambers and express the zeroth line bundle cohomology as an index. These chambers were either Mori chambers (K\"ahler cones of birational models of $X$) or Zariski chambers. In many cases, there are symmetries relating the cohomology dimensions between different Mori chambers. On the manifold 7863 the number of Mori chambers turned out to be infinite. These results are the first steps towards a general prescription or possibly ``master formula" which allows deriving analytical cohomology formulae for three-folds, in terms of basic topological quantities such as the intersection numbers.

Some words of caution are in order. Most of the results presented here are grounded in the calculation of a small number of line bundle cohomology dimensions and hence have a conjectural character. However, we have performed many non-trivial checks which indicate that the general picture is correct. 

Out of the 36 Picard number 2 families of CICY three-folds, in 34 cases the above techniques allow a description of all zeroth cohomologies. This includes the bicubic (the Calabi-Yau hypersurface of degree $(3,3)$ in $\IP^2\times \IP^2$), which is particularly simple as in this case the effective and the nef cones overlap. On the remaining 2 families of manifolds, with identifiers 7821 and 7809, there are computational difficulties in finding a sufficient number of cohomology values to facilitate the analysis.

It remains to be seen how much of the present discussion can be generalised to other Calabi-Yau three-folds, including examples with larger Picard number. There are preliminary indications that similar structures can be found in CICY three-folds with Picard number $3$. Exploring higher cohomology is another important topic to which we would like to return in a future publication. 

\section*{Acknowledgements}
We are grateful to Rehan Deen and Fabian Ruehle for discussions. AC would like to thank EPSRC for grant $\text{EP/T016280/1}$.


\providecommand{\href}[2]{#2}\begingroup\raggedright\endgroup

\end{document}